ARTICLE TEMPLATE

# A Novel Multilevel Taxonomical Approach for Describing High-Dimensional Unlabeled Movement Data

Yashar Tavakoli[a] and Lourdes Pena-Castillo[a,b] and Amilcar Soares[c]

[a]Department of Computer Science, Memorial University of Newfoundland, St. John's, NL, Canada;
[b]Department of Biology, Memorial University of Newfoundland, St. John's, NL, Canada;
[c]Deparment of Computer Science and Media Technology, Linnaeus University, Växjö, Sweden.

**ABSTRACT**

Movement data is prevalent across various applications and scientific fields, often characterized by its massive scale and complexity. Exploratory Data Analysis (EDA) plays a crucial role in summarizing and describing such data, enabling researchers to generate insights and support scientific hypotheses. Despite its importance, traditional EDA practices face limitations when applied to high-dimensional, unlabeled movement data. The complexity and multi-faceted nature of this type of data require more advanced methods that go beyond the capabilities of current EDA techniques. This study addresses the gap in current EDA practices by proposing a novel approach that leverages movement variable taxonomies and outlier detection. We hypothesize that organizing movement features into a taxonomy, and applying anomaly detection to combinations of taxonomic nodes, can reveal meaningful patterns and lead to more interpretable descriptions of the data. To test this hypothesis, we introduce TUMD, a new method that integrates movement taxonomies with outlier detection to enhance data analysis and interpretation. TUMD was evaluated across four diverse datasets of moving objects using fixed parameter values. Its effectiveness was assessed through two passes: the first pass categorized the majority of movement patterns as Kinematic, Geometric, or Hybrid for all datasets, while the second pass refined these behaviors into more specific categories such as Speed, Acceleration, or Indentation. TUMD met the effectiveness criteria in three datasets, demonstrating its ability to describe and refine movement behaviors. The results confirmed our hypothesis, showing that the combination of movement taxonomies and anomaly detection successfully uncovers meaningful and interpretable patterns within high-dimensional, unlabeled movement data.

Email: yt3165@mun.ca
Email: lourdes@mun.ca
Email: amilcar.soares@lnu.se

## 1. Introduction

Movement data is central to applications and scientific research involving various moving objects, natural (e.g., hurricanes, animals, icebergs, asteroids) or human-made (e.g., cars, ships, planes, buoys). Movement data is complex and massive (Dodge *et al.* 2008) and making the use of Exploratory Data Analysis (EDA) in the context of hypothesis generation is a well-established practice of *describing* the behavior of moving objects entailed by labeled or unlabeled data (Andrienko and Andrienko 2006, Andrienko *et al.* 2013a, Ma *et al.* 2017). The increasing interest in eXplainable AI (XAI) methods highlights the importance of transparency and interpretability in data analysis. Like XAI, EDA methods aim to build trust and provide deeper insights into the data, making them crucial for understanding and interpreting complex movement behaviors. Therefore, providing clear and interpretable descriptions of movement data enhances the usability and reliability of AI tools and techniques applied in various domains (Linardatos *et al.* 2020, Samek 2019, Kamath 2021, Thampi 2022, Samek and Müller 2019, Dosilovic *et al.* 2018, Abreu *et al.* 2021). We also face the rapid emergence of Large Language Models (LLMs, e.g., GPT, Gemini, BLOOM, etc.) and Recommendation Systems. "Describing" data is essential to LLMs and Recommendation Systems (Acharya *et al.* 2023, Alaaeldin *et al.* 2021). Such technologies are of interest in Geospatial Data Analysis as well (Jiang and Yang 2024), which opens up the possibility of structured data, and in particular tabular data in LLMs and Recommendation Systems revolving around Geospatial Data (Sui *et al.* 2024), which is a direction which the current study is facilitating.

This study proposes a novel EDA method to describe unlabeled high-dimensional movement datasets. While a method exists for labeled movement data (Tavakoli *et al.* 2022), no method exists for unlabeled movement data. This is a noteworthy shortcoming for multiple reasons: (i) unlabeled movement data (just like labeled movement data) is often high-dimensional (Staff 2016, Sturm *et al.* 2019, Tavakoli *et al.* 2022); (ii) the development of scientific hypotheses heavily relies upon data description (labeled or unlabeled) (Grolemund and Wickham 2014, Buja *et al.* 2009, Bergman 2014, Jebb *et al.* 2017, Ho Yu 2010, Fife and Rodgers 2022, Cuadrado-Gallego and Demchenko 2020, Behrens 1997, Good 1983, Abt 1987) and this is, in particular, evident in scientific fields involving moving objects (Andrienko *et al.* 2013b, Renso *et al.* 2013); (iii) unlabeled data, in general, is less expensive and abundantly available compared to labeled data (Chu *et al.* 2012, Mahdavifar *et al.* 2020, Denis *et al.* 2005, De Comité *et al.* 1999, He *et al.* 2022, Wang *et al.* 2019a, Park *et al.* 2003, Soares Junior *et al.* 2017); finally, (iv) one can argue that labels in data, in general, can describe the data (albeit minimally) through the preconceived traits of each class. Unlike data descriptions for labeled data, data descriptions for unlabeled data are generally more vague due to the absence of prior knowledge embedded in the labels. Labels provide a baseline understanding and context that guide the interpretation of the data. Without labels, validating and contextualizing data descriptions becomes more challenging because there is no predefined knowledge that typically guides the analysis. Therefore, a data description method for unlabeled movement data would be even more helpful than its counterpart for labeled data, and we propose a novel method to this end.

In this study, our hypothesis is that by evaluating the features through a carefully designed movement taxonomy and splitting the feature set into more interpretable higher-level representative nodes in this taxonomy, we can uncover meaningful patterns that manifest varying degrees of "interestingness" with respect to this taxonomy. The underlying premise is that breaking down the high-dimensional data into coherent behavioral groups allows for a more focused analysis of movement behaviors, making it easier to detect rare or unexpected patterns that may otherwise be obscured in the raw high-dimensional data. If such patterns can be found by such an strategy, we enhance the interpretability of the movement behaviors and increase the likelihood of identifying anomalies that are not immediately apparent by aligning the features with the structure provided by the taxonomy. We propose method to this end that comprises two key steps: the use of *movement taxonomies* and *anomaly detection* strategies. In the first step, a movement taxonomy aggregates variables (i.e., feature sets) into coherent movement behaviors defined by the taxonomy's nodes and leaves, reducing the data dimensionality. In the second step, an anomaly detection method measures the behavior of each data instance (i.e., moving object) regarding the designated coherent patterns determined in the chosen taxonomy. In this work, the anomaly detection method is used to measure the degree of interestingness of a movement pattern. In the context of our work, anomalies represent rare, unexpected, and potentially significant behaviors that stand out from the majority of the data. These characteristics make anomalies inherently interesting for further investigation and analysis. In this work, we rely upon a simple yet practical taxonomy of movement behaviors and four unlabeled movement datasets of diverse natures to evaluate the proposed method. We first select two nodes from the taxonomy (e.g., Nodes X and Y) and apply the anomaly detection method for each group of features belonging to these nodes. We then categorize each movement data instance into four zones. The *Zone 0* (i.e., manifests a common behavior) characterizes movements that do not manifest any of the two compared feature sets from the proposed taxonomy. *Zones 1* (i.e., manifests an uncommon behavior for element Y) and *2* (i.e., manifests an uncommon behavior for element X) characterize the movements manifesting a single feature set given by the evaluated taxonomy. Finally, *Zone 3* (i.e., hybrid) characterizes the movements that manifest a hybrid behavior that can not be attributed solely to a single feature set. The evaluation of our method involved characterizing the movement behavior of the majority of the data instances that were categorized under the four proposed zones. Our results support our hypothesis that by organizing features into a movement taxonomy and applying anomaly detection methods to combinations of elements of the taxonomy, we can uncover meaningful patterns of interest within high-dimensional, unlabeled movement data. We demonstrate in this paper that our method not only simplifies the complex, high-dimensional nature of the movement data but also makes the process of identifying significant patterns more intuitive and insightful. The detection of anomalous behaviors, those that deviate from the common patterns represented by the taxonomy's nodes, provides valuable insights into movements that are rare, surprising, or potentially impactful. These behaviors, which fall into what we categorize as Zones 1, 2, and 3, often represent the most interesting and important findings, as they highlight behaviors that differ meaningfully from the expected or majority patterns, offering avenues for deeper exploration and understanding of movement data.

This work is organized as follows. Section 2 presents a literature review. Section 3 introduces our proposed method to describe unlabeled movement data. Section 4 details how the proposed method is evaluated. The evaluation results are presented in Section 5. Finally, Section 6 summarizes the present article and discusses some ideas as to how the proposed method can be improved and expanded.

## 2. Literature review

In its native form, each observation from movement data is a sequence of triplets $(x_1, y_1, t_1), (x_2, y_2, t_2), \ldots, (x_n, y_n, t_n)$ where $n \geq 1$ is an integer (i.e., an index for the data sequentially collected from the moving object). Variables $x_i$ and $y_i$ are the spatial location that the moving object has visited at step $i$ (where $i$ is an integer and $1 \leq i \leq n$), and $t_i$ is the time measurement when the moving object has visited the locations $x_i$ and $y_i$. As such, a sequence of triplets, also known as a trajectory, signifies all the locations a moving object has visited ordered through time (Renso et al. 2013). Therefore, movement data in its native form is a collection of trajectories.

Raw trajectories are not comprehensible entities per se, albeit usable by machine (Rintoul and Wilson 2015, Wilson et al. 2016). Comprehensible forms of movement data, however, do exist. For example, a form of comprehensible movement data derived from trajectories, which we call *movement parameter* here, is a sequence of scalars portraying a trait of a moving object through time (e.g., Speed, Acceleration, turn angles) (Dodge et al. 2009). In this form, movement data consists of movement parameter(s) across all the trajectories. Single or multiple statistical variables (e.g., mean, variance, percentiles, skewness) can represent the distribution of a movement parameter for a given trajectory (Dodge et al. 2009, Haidri et al. 2022). Our study refers to these variables as *movement variables*. Beyond just representing the distribution of a movement parameter, a descriptor can directly summarize an aspect (e.g., straightness, length, duration, perimeter/area of the associated convex-hull) of a trajectory with a scalar (Rintoul and Wilson 2015). In this form, movement data consists of a collection of vectors (each representing a trajectory). Variable-based movement data has proved to be a highly comprehensible form of movement data, as movement variables summarize a trajectory into a vector (Laube 2014, Rintoul et al. 2021, Soleymani et al. 2014).

According to Andrienko and Andrienko (2006), EDA over movement data studies distributions of (and subsequent variations or trends in) movement variables to describe the behavior of moving objects. Basically, the value of a movement variable over all the trajectories in a cluster constitutes a distribution (Andrienko and Andrienko 2006). Researchers rely upon these distributions to understand the behavior of moving objects. When movement data is unlabeled, the reference of EDA over movement data is called *clusters* of moving objects based on movement variables (Laube 2014), and the subsequent study of distributions of movement variables across all the underlying clusters. Summary statistics, in general, have proved to be a convenient tool in studying the nuances imparted by a distribution (Pearson 2018, Martinez et al. 2010). With movement data in particular, some scholars prefer to present summary statistics with different box plots, while others prefer to table the raw magnitudes (Cachat et al. 2011, Sheng and Yin 2018, Xu et al. 2021, Guo et al. 2020, Kontopoulos et al. 2020, Sharma et al. 2020, Pornsupikul et al. 2017).

In all previously mentioned works, researchers restrain their experiments to utilizing a limited set of movement variables to conduct EDA. The reason is that conducting EDA essentially becomes very difficult in higher dimensions (Waggoner 2021, Huang et al. 2017, Khattree 2015, Lespinats et al. 2015, Xia et al. 2018). However, a limited number of variables often fails to capture the full complexity of a trajectory, given that movement data is inherently high-dimensional. A trajectory contains extensive information, and the inclusion of more variables in a vectorized trajectory enhances its ability to accurately represent the target trajectory (Xiao et al. 2017, Staff 2016, Sturm et al. 2019, Tavakoli et al. 2022).

Dimensionality reduction (i.e., the transformation of a large set of variables into a smaller set) is the general technique that is widely used for EDA in cases where data is high-dimensional (Burges 2010, Wang 2012, Lanoiselée and Grebenkov 2017, Khattree 2015, Lespinats et al. 2015, Xia et al. 2018). Although dimensionality reduction techniques can uncover the structure of high-dimensional data, they transform the original variables (Bolón-Canedo et al. 2015, Kuhn 2019). However, intact movement variables are fundamental for data description due to their interpretability (Dodge et al. 2009, Rintoul and Wilson 2015). Consequently, dimensionality reduction may not fully or partially facilitate a comprehensive understanding of movement data.

Despite significant advances in EDA utilizing movement variables to summarize trajectories, existing literature predominantly focuses on low-dimensional datasets and labeled data (Soleymani et al. 2014, Laube and Purves 2011, Stein et al. 2017, Collares et al. 2018, Buzin et al. 2019, Stuart and Long 2011, Budge and Long 2018, Lee et al. 2021, Huang et al. 2020, Feng et al. 2021, Matzner et al. 2015, Pineda and Naval 2020, GUO et al. 2020, Wang et al. 2018, Kim and Lee 2020, Sheng et al. 2018, Etemad et al. 2018, Xiao et al. 2017, Chan et al. 2021, Bolbol et al. 2012, Golshan 2020, Xiao et al. 2017, Li 2014, Lashkari et al. 2018, Colefax et al. 2020, Da Silveira et al. 2016, Pandey and Liou 2020, Terry and Gienko 2011, Terry and Kim 2015, Terry and Feng 2010, Terry and Gienko 2018, Terry and Feng 2010, Zhang et al. 2018). The current methodologies for high-dimensional data, which often involve dimensionality reduction techniques, compromise the interpretability of movement variables, thus limiting the comprehensiveness of the analysis. While (Tavakoli et al. 2022) has attempted to address the complexity of high-dimensional movement data through taxonomies, it necessitates labeled data; furthermore, it fails to leverage the full potential of multilevel taxonomies for a more granular understanding. Therefore, a critical gap exists in effectively describing high-dimensional, unlabeled movement data without transforming the original variables. Our work aims to bridge this gap by extending the single-level taxonomical description of labeled movement data proposed by Tavakoli et al. (2022) to a multilevel taxonomical description of unlabeled data. This novel approach enables a more profound and finer description of movement behaviors in unlabeled datasets, enhancing the comprehensibility of movement data analysis.

## 3. A multilevel taxonomical approach for describing unlabeled movement data

In its initial form, movement data consists of an incomprehensible collection of trajectories, making it challenging to discern the underlying movement behaviors. This challenge is particularly prominent when dealing with unlabeled data. However, the inherent properties of movement variables offer a potential solution: these variables can be systematically organized into *taxonomies* (Wang et al. 2019b, Kontopoulos et al. 2020, Fu et al. 2017, Beyan and Fisher 2013, Yang et al. 2018, Laube 2014, Wiratma et al. 2017). By employing a taxonomy, movement variables can be categorized into distinct, non-overlapping nodes/categories. For instance, a straightforward taxonomy might consist of two categories: *Kinematic* and *Geometric*. In this taxonomy, "Speed" would be placed under the Kinematic category, while "curvature" would fall under the Geometric category. This structured approach facilitates a more comprehensible analysis of movement data, enabling practitioners to better interpret and understand the behaviors encapsulated within the trajectories.

The taxonomical representation of an entity or a phenomenon helps its cognition by hierarchically representing its structure (Pavlinov 2021). For example, represent-ing movement variables through the Kinematic/Geometric taxonomy imparts to the practitioner that a subset of variables concerns the *motion* of a moving object (i.e., Kinematic variables). In contrast, another subset concerns the *shape* of the trajectory of moving objects (i.e., Geometric variables) where motion and shape are primitive concepts of movement behavior.

In this context, a data instance composed of Kinematic and Geometric variables may encapsulate interpretable information about a moving object's behavior. We observed that *outlier scores* obtained from anomaly detection methods can effectively describe such information. An *outlier score*, as formally defined, is a numerical value assigned to each data instance by an anomaly detection algorithm, representing the degree of deviation of that instance from the norm within the dataset. Higher outlier scores indicate a greater degree of abnormality or atypical behavior than the rest of the data. As noted by Suri *et al.* (2019), outliers can highlight the degree of *interestingness* or *significance* (to the analyst) of a data instance within a dataset, as it provides valuable insights into uncommon or atypical movement behaviors. Therefore, in this paper, when we refer to the degree of interestingness of a movement behavior, we mean how uncommon or atypical it is concerning the normal behavior present in the dataset. For instance, in the context of movement data, consider several moving objects with average Speeds around 1 $m/s$ and a single moving object with an average Speed of 10 $m/s$. This latter object can be considered significantly more remarkable regarding average Speed, as it deviates considerably from the typical movement behavior exhibited by the others. Similarly, in a higher-dimensional set of Kinematic variables, the magnitude of an outlier score for a given data instance reflects the extent to which it deviates from typical Kinematic behavior relative to other data instances. Given that a vectorized dataset effectively summarizes and represents a collection of trajectories and that taxonomies of variables enhance comprehension, it is possible to describe an unlabeled movement dataset by utilizing outlier scores for all (or selected) categories within a given taxonomy. Building on this rationale, we propose an approach to describe unlabeled movement data in its raw form, as illustrated in Figure 1. We call this approach TUMD (Taxonomical description of Unlabeled Movement Data) for simplicity. TUMD comprises four *artifacts* and three *processes*. An artifact here is any usable entity, and a process transforms a given artifact into another. Next, we detail TUMD by explaining each process.

Process 1 is informed by three key artifacts: a taxonomy, a dataset of trajectories, and a description from previous analysis. This process generates a vectorized dataset based on movement variables derived from selected taxonomy nodes. The variables are primarily determined by the nodes chosen from the taxonomy without the necessity of including all nodes. Additionally, the dataset description from previous analyses may guide the selection of movement variables, mainly if it provides a rationale for investigating other taxonomy nodes that have not yet been studied.

Next, Process 2 is informed by two key artifacts: the vectorized dataset and the given taxonomy. Based on the vectorized dataset, this process generates a series of outlier scores for each selected data instance. If the analysis involves $k$ nodes from the taxonomy, Process 2 produces $k$ outlier scores for each selected data instance. The given taxonomy dictates which movement variables will be included in the computation of outlier scores for each node.

Finally, Process 3 is informed by two key artifacts: the outlier scores and the given taxonomy. This process generates a description of the vectorized dataset, which summarizes and represents the original dataset of trajectories. The magnitude of an outlier score indicates whether a data instance exhibits the movement behavior associated with the corresponding taxonomy node. A verbal summary of these indications constitutes the target description.

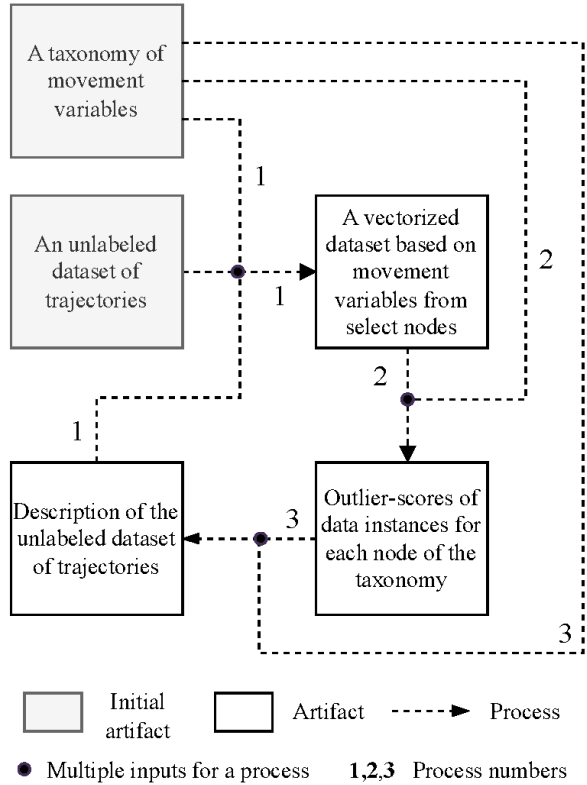

Figure 1.: The flowchart of our proposed approach (TUMD) to describe unlabeled movement data. TUMD is iterative in that it can accommodate multiple passes.

TUMD is designed to be generic, supporting various methods based on several technical decisions left open as parameters. The parameters of TUMD are detailed in Table 1.

## 4. Evaluation scheme

To evaluate TUMD, we focused on two main areas: parameter choices and effectiveness measures. The evaluation involves selecting appropriate parameter settings, for which we have opted for simple yet practical choices to streamline the process. This subsection details the methodology for conducting a data-oriented evaluation of TUMD by addressing these two focal topics.

### 4.1. *Movement Parameters and Taxonomy*

The first parameter of TUMD is the selection of a taxonomy. We utilize the taxonomy illustrated in Figure 2, which comprises two levels. At the first level, movement variables are categorized into two groups: *Geometric*, which includes variables related to the shape of the trajectory, and *Kinematic*, which includes variables related to the

Table 1.: Parameters' descriptions for TUMD

| Parameter | Description |
| --- | --- |
| A taxonomy of movement variables | A taxonomy of movement variables is required by all the three processes in TUMD. The taxonomy can be as simple as a single-level two-node tree or more complicated with many levels and nodes. |
| Movement variables | Each taxonomy node must include at least one movement variable. |
| Outlier scoring technique | There are many techniques for outlier scoring to choose from Boukerche *et al.* (2020), based on which Process 2 may produce outlier scores. |
| Decision boundaries to determine movement behaviors | If the magnitude of an outlier score corresponding to a node is large for a moving object, then Process 3 decides that the object manifests the behavior captured by the node. Therefore, each node must have a decision boundary defining "large" (or small). |
| Start and feedback | It must be specified which nodes of a given taxonomy are subject to TUMD in the beginning; and how the most current produced description can inform Process 1 as to what nodes of the taxonomy are to be considered next. |
| Summarization tool | Numerous data instances result in many outlier scores. A summarization tool (such as visualization and/or summary statistics) is required in Process 3 to provide a comprehensible basis for a verbal description of the movement behaviors latent in a dataset. |

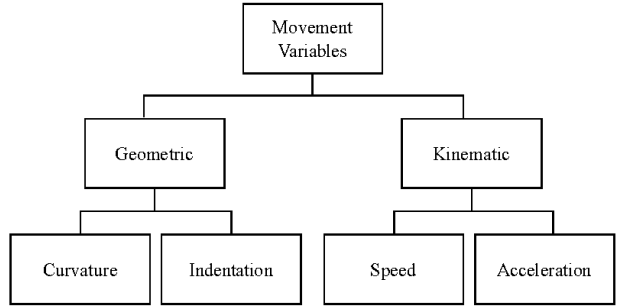

Figure 2.: The taxonomy of movement variables we use to evaluate TUMD.

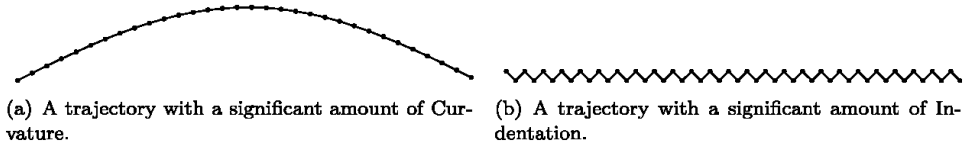

(a) A trajectory with a significant amount of Curvature.

(b) A trajectory with a significant amount of Indentation.

Figure 3.: Difference between Curvature and Indentation in trajectories. Note that the dots on the trajectories indicate the sequence of locations that the underlying moving objects have visited.

moving object's motion.

At the second level, Geometric variables are subdivided into *Curvature* and *Indentation* (Figure 3). As Figure 3(a) shows, the Curvature of a trajectory is the amount to which a trajectory deviates from a straight line. Indentation, on the other hand, as depicted by Figure 3(b), is the amount of course changes at each point of the trajectory. It must be noticed that, as portrayed in Figure 3, a trajectory with a significant Curvature may not manifest a substantial amount of Indentation. Likewise, a trajectory with significant Indentation may not manifest substantial Curvature.

At the second level of the taxonomy, the Kinematic node is divided into two categories: *Speed* and *Acceleration*. Speed includes all movement variables related to the distance covered over time, while Acceleration encompasses variables associated with the rate of Speed change over time. The proposed taxonomy for evaluating TUMD prioritizes simplicity and practicality. This taxonomy has a minimal number of nodes, each having two child categories, allowing the outlier scores to be expressed in two dimensions. This structure simplifies parameter choices for TUMD while maintaining an intuitive framework that facilitates the comprehension of movement datasets.

Table 2 lists 72 widely used variables from various scientific disciplines incorporated in the taxonomy to evaluate TUMD. Several statistical variables are utilized for the Indentation, Speed, and Acceleration nodes due to their prevalent use in the literature for measuring movement parameters. For measuring the Curvature of a trajectory, which reflects its overall shape, Distance Geometries is the most effective tool (Rintoul and Wilson 2015).

It is crucial to note that movement parameters, such as Acceleration, Speed, Indentation, and Curvature, can each be measured and represented by a single variable (e.g., average values). However, movement parameters (and thus movement data) are inherently high-dimensional, and a single value does not capture the full range of a movement parameter's manifestation. Multiple variables are used to measure the movement parameters (Speed, Acceleration, Curvature, and Indentation) within the Geometric and Kinematic categories to describe movement behavior comprehensively. This approach ensures a more complete representation of the movement data, capturing the complexity and nuances of the movement behavior.

### 4.2. *Outlier scoring method and movement behavior zone categorization*

The outlier scoring technique is another crucial parameter in TUMD. Numerous techniques exist for outlier scoring, ranging from simple to complex methods (Boukerche et al. 2020). We utilize *Distance-based Outlier Detection* (Knorr and Ng 1997) to maintain simplicity and avoid unnecessary complexity. The core idea of the Distance-based Outlier Detection algorithm is to determine whether a data point is an anomaly by examining the number of other points within a certain distance (radius) from it. Specifically, a point is considered an anomaly if it has fewer than a specified number of neighboring points within this distance. To quantify this, the method assigns a score between 0 and 1 to each data instance based on the number of neighboring instances. A score close to 0 indicates that a data instance is less of an anomaly (i.e., a more common movement pattern). In contrast, a score close to 1 indicates it is more of an anomaly (i.e., a more interesting movement pattern). More accurately, in all our experiments, the outlier scores are calculated based on the number of data instances that fall within the expected radius in a given dataset: for a data point, a score of 1 is assigned if no point falls within the expected radius; a score of 0 is assigned if all the points fall within the expected radius; in other cases, a score between 0 and 1 is assigned accordingly in a uniform fashion. In this work, we used a fixed radius value as the average distance between all data instances in a given dataset. The average distance between the instances represents the overall distribution of the data points. This value reflects the typical spacing between data points, ensuring that the method is neither too lenient nor too strict in identifying outliers while minimizing potential bias that could arise from arbitrary radius selection.

With the outlier scores standardized between 0 and 1, we can define decision boundaries to categorize movement behaviors. Given that our target taxonomy expresses outlier scores in two dimensions, we can also express decision boundaries in two dimensions. We propose decision boundaries as illustrated in Figure 4 to classify the movement patterns. To explain Figure 4, first note that each taxonomy node consists of coherent variables encapsulating movement behavior. Now assume two groups of variables (belonging to two separate nodes in the chosen taxonomy) denoted by $X$ and $Y$ represented on the horizontal and vertical axes of Figure 4, respectively. A fundamental analysis might use the diagonal as a boundary to determine whether data instances exhibit $X$ behavior more than $Y$ or vice versa. However, we introduce more expressive decision boundaries: assuming that outlier score values below 0.5 imply more common movement behaviors, four distinct zones emerge. The outlier score of a given data instance $d$ can hypothetically fall within any of these four zones:

- In *Zone 0*, which we call **common behavior**, where the outlier scores for both $X$ and $Y$ are below 0.5, $d$ manifests no uncommon $X$ or $Y$ behavior.
- In *Zone 1*, which we call **uncommon Y**, where the outlier score for $Y$ exceeds 0.5 while the score for $X$ is below 0.5, $d$ purely manifests uncommon behavior with regards to the $Y$ set of variables from the taxonomy.
- In *Zone 2*, which we call **uncommon X**, where the outlier score for $X$ exceeds 0.5 while the score for $Y$ is below 0.5, $d$ purely manifests uncommon behavior with regards to the $X$ set of variables from the taxonomy.

Table 2.: Movement variables used for the evaluation of TUMD

| Leaf | Movement variables | Description |
|---|---|---|
| Curvature | 5 signatures of Distance Geometries consisting of 15 variables. | 15 variables that measure straightness through effective distance (the ratio of the distance between the start and end points of a segment to the length of the segment where a value of 1 indicates a completely straight trajectory and a value close to 0 indicates a tortuous one). Each summation term, called a *signature*, measures the straightness in progressively finer frequencies. Therefore, the first signature is one variable and is the effective length of the entire trajectory; the second signature consists of two variables (the first being the effective length of the first segment and the second being the effective distance of the second segment), and subsequently. |
| Indentation | 19 statistical variables: unique values, number of 0's, arithmetic mean, standard error of the mean, quantiles, standard deviation, coefficient of variation, median absolute deviation, interquartile range, skewness, kurtosis, and the lowest and the highest values. | The statistical variables summarize the distribution of the angles at each point visited by a moving object over its entire movement. |
| Speed | 19 statistical variables as listed for Indentation. | The variables summarize the distribution of the Speed magnitude measured at each point over its entire movement. |
| Acceleration | 19 statistical variables as listed for Indentation. | The variables summarize the distribution of the Acceleration magnitude measured at each point over its entire movement. |

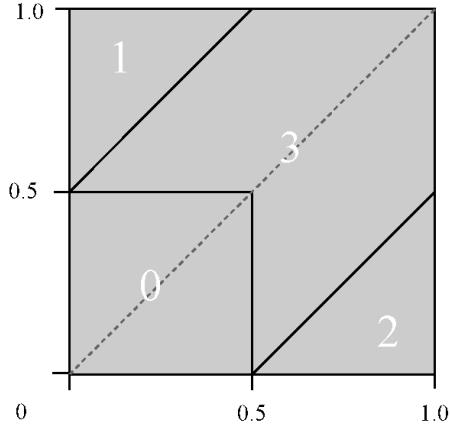

Figure 4.: Decision boundaries to determine movement behaviors consisting of two categories in movement taxonomy.

- Finally, in *Zone 3*, the movement behavior of $d$ is **hybrid**; although the behavior is discernible, it cannot be attributed solely to uncommon behavior with regards to the $X$ or $Y$ set of variables from the taxonomy.

This structured approach allows for a nuanced understanding of movement behaviors based on outlier scores, providing clear and interpretable insights into the data.

The start and feedback parameters, which are critical for evaluating the second pass of TUMD, are discussed below. The evaluation begins at the first level of the target taxonomy, producing outlier scores for certain nodes. Data instances (i.e., movement patterns) that exhibit pure Geometric behavior (Zone 1, with Geometric on the vertical axis in Figure 4) undergo another iteration of TUMD to determine whether their Geometric character can be attributed to Curvature or Indentation. Similarly, data instances that exhibit pure Kinematic behavior (Zone 2, with Kinematic on the horizontal axis in Figure 4) undergo another iteration of TUMD to ascertain whether their Kinematic character is due to Acceleration or Speed. This practical iterative approach allows the practitioner to start with a general movement behavior and progressively narrow it down to more specific characteristics.

Finally, the final parameter to discuss is the summarization tool, which helps determine the movement behavior of multiple data instances involving various outlier scores. Each iteration of TUMD evaluation generates outlier scores in two dimensions, making scatter plots a practical choice for summarization. Additionally, we present the results across all experiments using donut charts, which comprehensively visualize the movement behaviors identified. We also use plots with the raw trajectories and density plots to assess the patterns assigned to the proposed four zones.

### 4.3. *Effectiveness measure*

This study evaluates TUMD by examining a realization of it with the parameter choices detailed in Section 4.1. A successful application of TUMD in the first pass should effectively describe the movement behaviors hidden in a movement dataset. The first pass of TUMD is considered effective if the majority of the data instances (more than 50%) fall outside the "uninformative" Zone 0 (as shown in Figure 4), indicating that Kinematic, Geometric, or hybrid behaviors can be attributed to most data instances.

The second pass of TUMD, on the other hand, is defined as effective in the case of Geometric data instances if cumulatively more than 50% of the data instances determined as Geometric (in the first pass) fall within Zone 1 (Indentation) or Zone 2 (Curvature). This means that the second pass of TUMD has been able to further attribute Geometric behaviors of the majority of data instances to either uncommon Indentation or Curvature. Likewise, the second pass of TUMD is defined to be successful in the case of Kinematic data instances if cumulatively more than 50% of the data instances determined as Kinematic (in the first pass) fall within Zone 1 (Acceleration) or Zone 2 (Speed). The second pass of TUMD is defined to be successful in the case of Geometric data instances if cumulatively more than 50% of the data instances determined as Geometric (in the first pass) fall within Zone 1 (Indentation) or Zone 2 (Cruvature). Now, we call the second pass effective if it is successful for both Geometric and Kinematic data instances or either of them; otherwise, we call it ineffective. Both ineffective passes imply that TUMD, with the chosen parameters, cannot find uncommon movement behavior among most data instances to effectively describe the dataset. The definition of the introduced effectiveness measure guarantees that for at least some datasets, TUMD can detect the main behaviors of (and accordingly describe) the majority of underlying data instances (in the first pass) and then refine the description of at least some of the detected behaviors.

The evaluation scheme significantly depends on the characteristics of the underlying dataset. There might be limited distinctive Geometric or Kinematic behavior in datasets of specific sizes and natures (referring to the type of the underlying moving objects). To address this, we have diversified the target datasets to encompass a broad spectrum of possible and realistic movement datasets of varying sizes and origins, including natural and human-made objects.

We consider TUMD effective if, for at least one dataset under study, both the first and second passes meet their respective criteria for effectiveness. If no dataset satisfies these conditions, then TUMD is deemed ineffective for the chosen parameter combination. It is important to note that TUMD may still prove effective for the same datasets with a different set of parameters.

### 4.4. *Datasets*

In this study, we use four datasets of varied nature and size. The first dataset consists of 500 ship trajectories in the Pacific regions of Mexico, the United States, and Canada (delineated by a latitude between -180 and -110 and a longitude between 10 and 70) during the 5-day window from June 26, 2019, to June 30, 2019. All the trajectories are available as Vessel Traffic Data by Marine Cadastre[1]. The Automatic Identification System (AIS), the marine industry's most commonly used position reporting system, generates these trajectories.

The second dataset involves the movement of Arctic Foxes on Bylot Island (Nunavut, Canada), recorded during the sea ice period between July 2007 and May 2013, using Argos satellite telemetry data (Lai *et al.* 2016). This dataset includes the movement patterns of 66 foxes. According to (Lai *et al.* 2016), the primary objective of this dataset is to study the ecological processes shaping the ecosystem and to

---

[1]https://marinecadastre.gov/ais/

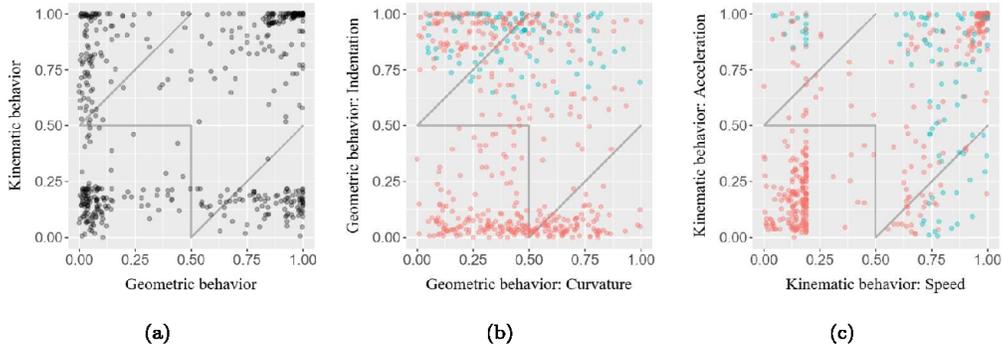

Figure 5.: Taxonomical description of the ships dataset. (a) Description of data instances based on the Geometric/Kinematic significance of their movement behavior (b) The significance of Curvature/Indentation in data instances with uncommon Geometric behavior ● in comparison with other data instances ● (c) The significance of Speed/Acceleration in data instances with uncommon Kinematic behavior ● in comparison with other data instances ●

understand the patterns of range residency, migration, and nomadism of the Arctic Fox.

The third dataset comprises 500 trajectories of tropical cyclones from 1842 to the present, provided by The International Best Track Archive for Climate Stewardship (IBTrACS) (Knapp *et al.* 2018). This dataset is compiled from various observing systems, including surface reports and satellite observations. As noted in (Knapp *et al.* 2018), studying tropical cyclone behavior is crucial in climatology due to its significant impact on human life, property, and ecology.

Our final dataset consists of trajectories from a 90-minute professional U19 football (soccer) match in Switzerland, available at [2]. This dataset includes the movement trajectories of 22 players from both teams. Previous research (Kim *et al.* 2011, Gudmundsson and Horton 2017, Stein *et al.* 2017) highlights that the descriptive analysis of sports, particularly football, has numerous applications, including performance analysis and strategy development.

## 5. Results

This section evaluates TUMD based on the evaluation scheme explained in Section 4 for each target dataset and then summarizes the results.

### 5.1. *Ships*

As indicated in Figure 5(a), 28% of the data instances fall within Zone 0, and 72% fall within the union of Zones 1, 2, and 3. This indicates that 72% of the ships exhibit uncommon movement behaviors. Specifically, 19% of the ships display pure Kinematic behavior (Zone 1), 17% exhibit pure Geometric behavior (Zone 2), and 36% show mixed Kinematic/Geometric behavior (Zone 3).

---

[2]https://old.datahub.io/dataset/magglingen2013

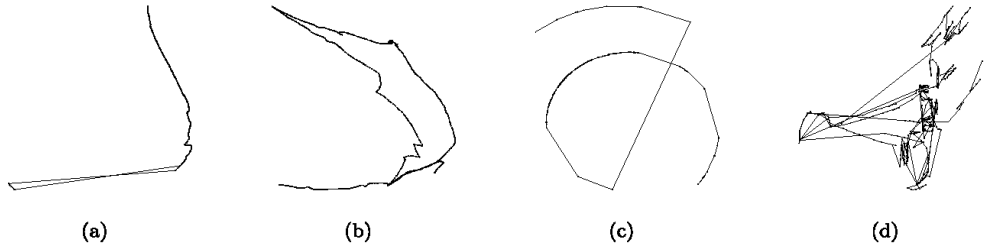

Figure 6.: Comparing the spatial component of a trajectory with uncommon hybrid Geometric behavior in (d) with three common ship trajectories in (a), (b), and (c).

Figure 5(b) illustrates the attribution of movement behavior to Curvature and/or Indentation for ship trajectories that exhibit pure Geometric behavior. Among these, 67% can be attributed solely to uncommon Indentation (with outlier scores of at least 0.65) (Zone 1). The remaining 33% of these ship trajectories exhibit hybrid Geometric behaviors, attributed to both uncommon Indentation and Curvature (Zone 3). Notably, some data instances, highlighted by • in Figure 5(b), exhibit high Curvature and/or Indentation outlier scores. This is expected, as these instances belong to Zone 3 in Figure 5(a), demonstrating both uncommon Geometric and Kinematic behaviors. Consequently, it is reasonable to anticipate high Curvature and/or Indentation scores for these instances in Figure 5(b). Figure 5(c) shows how the movement behavior of ships with pure Kinematic behavior can be attributed to Speed and/or Acceleration. For these ships, 16% exhibit pure Acceleration (Zone 1), and 27% exhibit pure Speed (Zone 2). Finally, 57% of the Kinematic behavior is attributed to both uncommon Speed and Acceleration (Zone 3).

Resorting to taxonomical categories, TUMD compactly describes the complex movement behavior of moving objects concealed in high-dimensional data. The details of such behavior, however, can be cumbersome, calling for further analysis of the underlying values of the variables. One convenient way of investigating the details of the behaviors is to visualize trajectories and/or their properties. Although the visualization might highlight some aspects latent in data, it may not reveal all the nuances captured by many variables.

Applying k-means clustering and the elbow method, we have clustered the ship trajectories in Zone 0 into three clusters and presented (the spatial components) of the three trajectories closest (based on Euclidean distance) to the three centers of the clusters as depicted in Figure 6. The mentioned clusters encapsulate the most common Geometric (and Kinematic) traits of the ships in the dataset. Such traits can be many and are reflected by the values of the variables. For example, in Figure 6, we have presented a ship trajectory which we call pattern $A$ belonging to Zone 3 of Figure 5(b). Therefore, pattern $A$ manifests uncommon Indentation and Curvature traits. In this case, visualization does manifest the uncommon characteristics of pattern $A$: Unlike those of the cluster centers, the shape of pattern $A$ is more complex, manifesting traveling back and forth to a point. Aside from the Curvature, the Indentation of pattern $A$ also looks different compared to traits in Zone 0, where the clusters manifest very slight indentation or indentation limited to a small segment of trajectories.

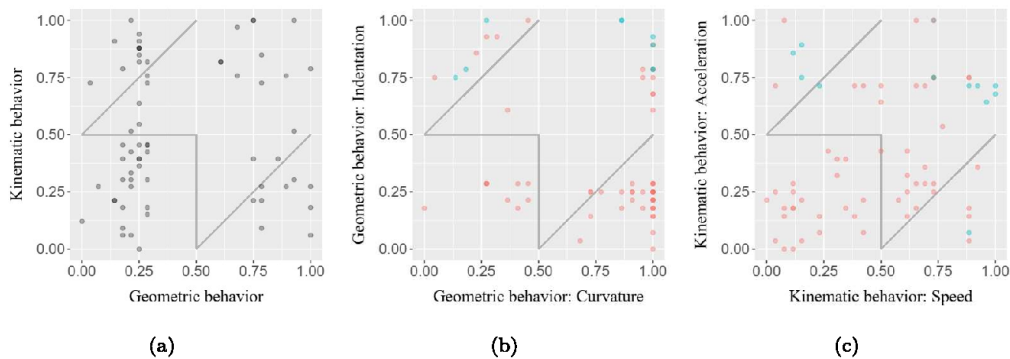

Figure 7.: Taxonomical description of the arctic foxes dataset. (a) Description of data instances based on the Geometric/Kinematic significance of their movement behavior (b) The significance of Curvature/Indentation in data instances with uncommon Geometric behavior • in comparison with other data instances • (c) The significance of Speed/Acceleration in data instances with uncommon Kinematic behavior • in comparison with other data instances •

## 5.2. *Arctic foxes*

The fox dataset is smaller than the ship dataset, and, unsurprisingly, the Distance-Based Outlier Detection algorithm results in fewer distinct outlier scores. This occurs because the set of distinct numbers of data instances in all possible neighborhoods is smaller, leading to a reduced number of distinct outlier scores. In Figure 7(a), 43% of the data instances fall within Zone 0, while 57% fall within the union of Zones 1, 2, and 3. This signifies that 57% of the Arctic foxes exhibit significant movement behaviors. Specifically, 21% of the foxes demonstrate pure Kinematic behavior (Zone 1), 12% display pure Geometric behavior (Zone 2), and 24% show mixed Kinematic/Geometric behavior (Zone 3).

Figure 7(b) illustrates how the movement behavior of fox trajectories with pure Geometric behavior can be attributed to Curvature and/or Indentation. Of these, 35% can be attributed solely to uncommon Indentation (with outlier scores of at least 0.65) (Zone 1), while the remaining 65% exhibit hybrid Geometric behaviors, attributed to both uncommon Indentation and Curvature (Zone 3).

Figure 7(c) shows how the movement behavior of foxes with pure Kinematic behavior can be attributed to Speed and/or Acceleration. Among these foxes, 35% exhibit pure Acceleration (Zone 1), and 5% exhibit pure Speed (Zone 2). Finally, 60% of the Kinematic behavior is attributed to uncommon Speed and Acceleration (Zone 3).

The two density plots in Figure 8 present the distribution of Speed measurements of two clusters in Zone 0. Therefore, the plots depict the common movement traits of arctic foxes as they relate to Speed. The density plot in Figure 8(c) belongs to a fox that we call pattern *B* here and whose data instance has emerged in Zone 2 of Figure 7(c). A side-by-side comparison demonstrates this fox's uncommon movement traits related to Speed. The shape of the density plot belonging to pattern *B* manifests discrepancies with regard to the clusters. Most notably, the density plot of pattern *B* is considerably less skewed than the other density plots.

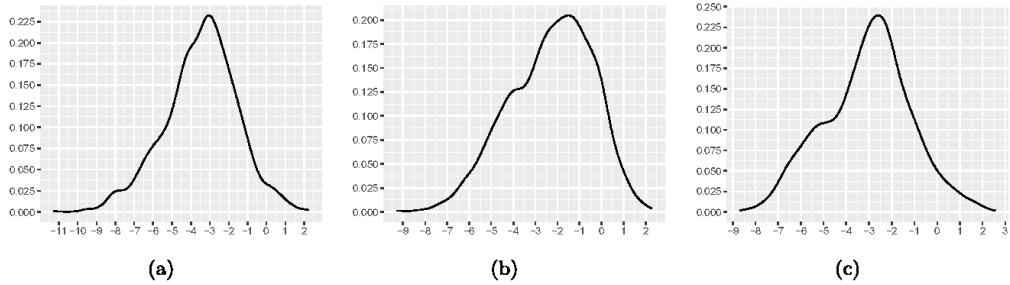

Figure 8.: Comparing the distribution of Speed measurements of a fox trajectory with atypical Speed-related Kinematic behavior in (c) with two typical distributions of Speed measurements of foxes (a) and (b). Vertical axes illustrate densities and horizontal axes illustrate Speed (after Power Transformation).

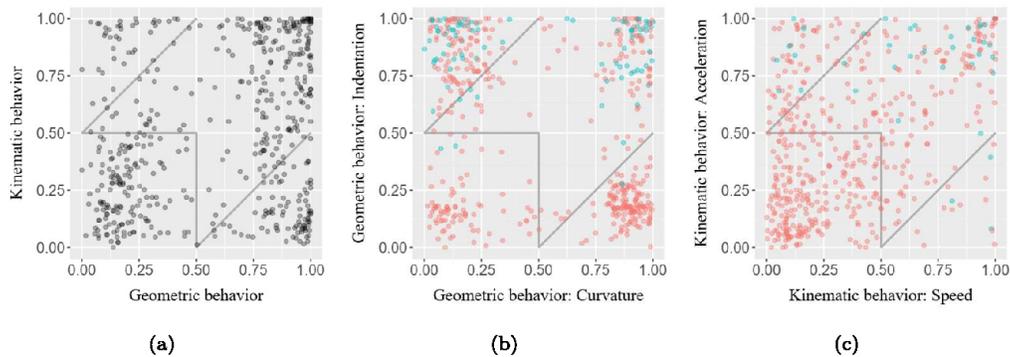

Figure 9.: Taxonomical description of the tropical cyclones dataset. (a) Description of data instances based on the Geometric/Kinematic significance of their movement behavior (b) The significance of Curvature/Indentation in data instances with uncommon Geometric behavior • in comparison with other data instances • (c) The significance of Speed/Acceleration in data instances with uncommon Kinematic behavior • in comparison with other data instances •

### 5.3. *Tropical cyclones*

In Figure 9(a), 29% of the data instances fall within Zone 0, while 71% fall within the union of Zones 1, 2, and 3. This indicates that 71% of tropical cyclones exhibit uncommon movement behaviors. Specifically, 1% of the cyclones display pure Kinematic behavior (Zone 1), 26% exhibit pure Geometric behavior (Zone 2), and 44% show mixed Kinematic/Geometric behavior (Zone 3).

Figure 9(b) illustrates how the movement behavior of cyclone trajectories with pure Geometric behavior can be attributed to Curvature and/or Indentation. Of these, 61% can be attributed solely to uncommon Indentation (Zone 1), while 38% exhibit hybrid Geometric behaviors, attributed to both uncommon Indentation and Curvature (Zone 3).

Figure 9(c) demonstrates how the movement behavior of cyclones with pure Kinematic behavior can be attributed to Speed and/or Acceleration. Among these cyclones, 41% exhibit pure Acceleration (Zone 1), and 6% exhibit pure Speed (Zone 2). Finally,

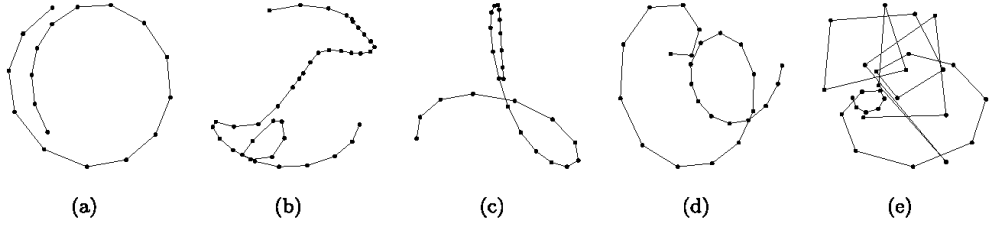

(a) (b) (c) (d) (e)

Figure 10.: Comparing the spatial component of a cyclone trajectory with uncommon Indentation-related Geometric behavior in (e) with four common cyclone trajectories in (a), (b), (c), and (d).

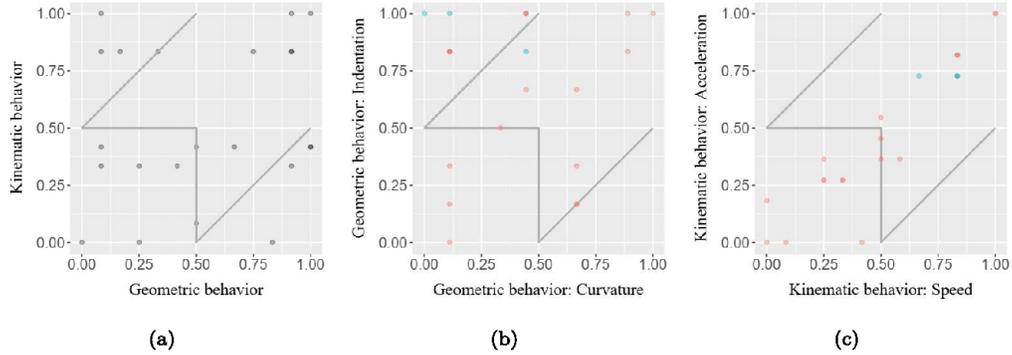

(a) (b) (c)

Figure 11.: Taxonomical description of the footballers' dataset. (a) Description of data instances based on the Geometric/Kinematic significance of their movement behavior (b) The significance of Curvature/Indentation in data instances with uncommon Geometric behavior ● in comparison with other data instances ● (c) The significance of Speed/Acceleration in data instances with uncommon Kinematic behavior ● in comparison with other data instances ●

53% of the Kinematic behavior is attributed to both uncommon Speed and Acceleration (Zone 3).

The four trajectories near the centers of clusters shown in Figure 10 (a, b, c, and d) illustrate the common Geometric traits of tropical cyclones observed in the dataset under study. The trajectory visualized in Figure 10(e), which we call pattern $C$, belongs to a data instance from Zone 1 in Figure 9(b). So, we expect that it manifests uncommon Geometric behavior related to Indentation traits. The visualization of the spatial components of this trajectory demonstrates this. Although the shape of pattern $C$ (measured by Curvature) is close to those in Figure 10(a) and Figure 10(b), its Indentation is different from pattern $C$ manifests many more dents. Note that low Indentation has emerged as a common Geometric trait of this dataset; in other datasets, other distribution-related traits (such as substantial Indetation) might emerge as common.

### 5.4. *Footballers*

In Figure 11(a), 27% of the data instances fall within Zone 0, while 73% fall within the union of Zones 1, 2, and 3. This indicates that 73% of the footballers exhibit uncommon movement behaviors. Specifically, 15% of the footballers display pure Kinematic behavior (Zone 1), 22% exhibit pure Geometric behavior (Zone 2), and 36% show mixed Kinematic/Geometric behavior (Zone 3).

Figure 11(b) illustrates how the movement behavior of footballers' trajectories with pure Geometric behavior can be attributed to Curvature and/or Indentation. Remarkably, 100% of these trajectories can be attributed solely to uncommon Indentation (Zone 1).

Figure 11(c) demonstrates how the movement behavior of footballers with pure Kinematic behavior can be attributed to Speed and/or Acceleration. Notably, the Kinematic behavior of 100% of these footballers can be attributed to both uncommon Speed and Acceleration (Zone 3).

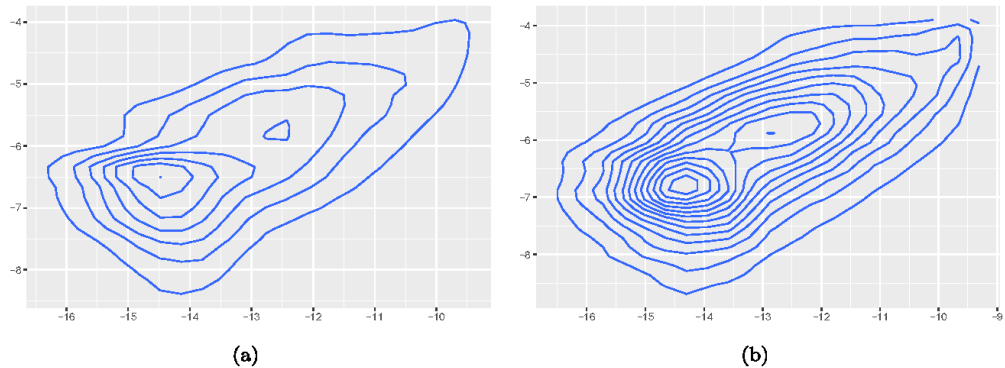

Figure 12.: Comparing the distribution of Speed/Acceleration measurements of a footballer trajectory with uncommon hybrid Kinematic behavior in (b) with the typical distribution of Speed/Acceleration of footballers (a).

The spatial component of trajectories belonging to footballers is so complex that their visualization cannot convey meaningful information about the uncommon Geometric traits of footballers as discovered and briefly described by TUMD as Indentation and/or Curvature. However, one can easily glimpse the uncommon Kinematic behavior of footballers as follows.

The density plot in Figure 12 (a) presents the two-dimensional distribution of Speed measurements against Acceleration of the center of Zone 0, depicting the common Kinematic traits of footballers. The density plot in Figure 12 (b), on the other hand, illustrates the Kinematic behavior of a footballer we call pattern $D$, belonging to Zone 3 of Figure11(c). Pattern $D$ manifests uncommon Kinematic traits, and in particular, we expect that pattern $D$ manifests uncommon hybrid Kinematic behavior. This is the case; the density plot shows that pattern $D$ manifests a sharper gradient over the Speed/Acceleration surface. Also, several segments of the contours belonging to pattern $D$ are convex, unlike what is common (Figure 12(a)). Moreover, pattern $D$ manifests joint Speed/Acceleration values, which are not common.

## 5.5. *Conclusion on the overall effectiveness*

We now evaluate the effectiveness of TUMD, referencing Figure 13. According to our criteria, TUMD is considered effective if it demonstrates effectiveness in both the first and second passes. We will first examine the effectiveness of TUMD in the first pass, followed by an assessment of its effectiveness in the second pass.

According to our definition of the effectiveness of the first pass, TUMD is effective if, at least for one data set, it can attribute the behavior of the majority of data instances as Kinematic, Geometric, or hybrid. Consequently, we showed that the first pass of TUMD is successful for all the datasets according to the definition, as in all datasets, less than 50% of data instances fall within Zone 0.

According to our definition of effectiveness in the second pass, the second pass is successful if the behavior of the majority of data instances recognized as either Kinematic or Geometric (or both) is refined further.

For the ships dataset, TUMD attributes the majority of Kinematic data instances (6%+4%=10% out of 19%) to either Speed or Acceleration, implying that the second pass of TUMD is effective on ship datasets. Even further, TUMD attributes the behavior of more than 50% (11% out of 17%) of the Geometric data instances to Indentation. For the foxes dataset, on the other hand, the second pass of TUMD is unable to refine either the Geometric behaviors or the Kinematic behavior of the majority of the data instances. For the cyclones dataset, TUMD attributes the majority of Geometric data instances (16% out of 26%) to Indentation. Therefore, the second pass of TUMD is effective for the cyclones dataset. For the footballers dataset, TUMD attributes the majority of Geometric data instances (22% out of 22%), to Indentation, which implies the effectiveness of the second pass of TUMD in this case as well.

Our definition of TUMD's effectiveness specifies that it is effective if, among the datasets under study, at least one dataset has both the first and second passes of TUMD effective. If TUMD does not meet this criterion with the chosen parameters, it is deemed ineffective. Consequently, we conclude that TUMD is effective, as the first pass is effective for all datasets, and the second pass is effective for the ships, cyclones, and footballers datasets.

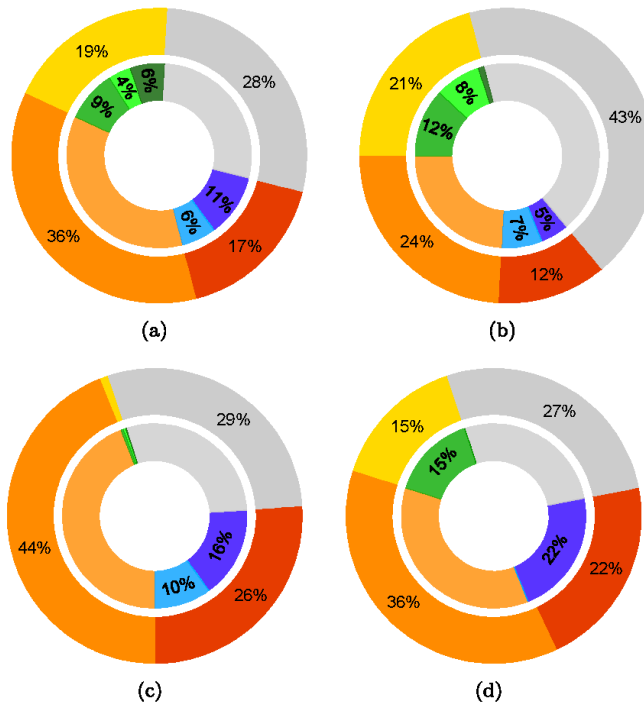

Figure 13.: A summary of the description produced by two levels of TUMD over ships, foxes, cyclones, and footballers datasets respectively depicted in (a), (b), (c), and (d). The outer ring captures a breakdown of the behaviors as they relate to the first level of the taxonomy, and the inner ring captures the second level. Color guide: Zone 0 (no discernible uncommon behavior) ■, Kinematic ■, Geometric ■, Hybrid Kinematic/Geometric ■, Speed ■, Acceleration ■, Hybrid Speed/Acceleration ■, Curvature ■, Indentation ■, Hybrid Curvature/Indentation ■.

## 6. Discussion

The results of this study, applied across diverse movement datasets, demonstrate that TUMD effectively describes unlabeled movement data in high-dimensional settings and specific parameter choices. However, alternative parameter selections could yield even better results, warranting further research into TUMD's parameters.

A key consideration is the taxonomy from a typology standpoint, which significantly influences TUMD's outcomes. Taxonomy levels can result from either intensional or extensional classification (Marradi 1990). Intensional classification covers all non-overlapping possibilities based on category definitions, such as distinguishing between Geometric and Kinematic movement variables. Conversely, extensional classification maximizes homogeneity within categories and heterogeneity between them, such as grouping Acceleration variables due to their strong internal homogeneity and distinctness from Speed variables. Recognizing overlaps or dependencies, such as higher Acceleration implying higher Speed, could refine TUMD to provide more accurate movement descriptions. To simplify the evaluation process, TUMD's decision bound-aries currently define four zones (Zone 0, Zone 1, Zone 2, and Zone 3). These boundaries, especially those in Zone 3, which capture hybrid movement behaviors, can be revised for better accuracy. Subdividing Zone 3 could yield more precise descriptions of hybrid behaviors. We introduced two passes in TUMD's evaluation process, focusing on start and feedback parameters. These passes can be modified or extended to introduce more granular evaluations. For instance, the second pass, which refines descriptions of pure Geometric and Kinematic behaviors, can be extended to further analyze sub-zones within Zone 3. While scatter plots were used for summarization, their effectiveness diminishes with large datasets or multiple categories. In such cases, alternative visualization techniques, sample-based scatter plots, or descriptive statistics might be more practical.

The authors of (Tavakoli et al. 2022) showed that a taxonomical approach could describe labeled data, which is particularly useful in XAI. However, the approach introduced in (Tavakoli et al. 2022) is limited to a single-level taxonomy and hence is open to the ideas of the present study to provide more accurate descriptions of movement behaviors with labeled movement data. Beyond descriptive analysis of movement behavior, TUMD has potential applications in feature selection. Knowledge of the feature space is crucial for effective feature selection in machine learning (Kuhn 2019, Liu. 1998, Duboue 2020, Huan Liu and Motoda 1998, Kumar 2018). TUMD, by providing detailed feature space insights, can enhance feature selection processes for movement data-related machine learning applications.

As discussed in Section 3, taxonomies are inherent to movement phenomena. It is conceivable that TUMD could be applied to other fields where taxonomies of variables exist or might be developed, following methodological practices as outlined by (Marradi 1990).

Finally, the effectiveness measure that we defined for TUMD entails its ability to describe the majority of data instances. So, a "collective" sense is implicit in the definition. However, TUMD can be useful as a tool for mining and analyzing individual data instances. This should be evident from the examples that this study proved with the visualization in the form of trajectories or distribution densities. In this regard, other effectiveness measures might be defined for TUMD as well.

## Funding

This research was partially funded by an Anonymous funding agency.

Andrienko, N. and Andrienko, G., 2006. *Exploratory analysis of spatial and temporal data: A systematic approach.* 1st ed. Berlin, Heidelberg: Springer-Verlag.

Behrens, J.T., 1997. Principles and procedures of exploratory data analysis. *Psychological Methods*, 2 (2), 131–160.

Bergman, J., 2014. Practical multivariate analysis, fifth edition. by a. afifi, s. may amp; v.a. clark. boca raton, florida: Chapman amp; hall/crc. 2011. 537 pages. uk£46.99 (hardback). isbn 978-1-4398-1680-6. *Australian amp; New Zealand Journal of Statistics*, 56 (4), 431–432.

Beyan, C. and Fisher, R., 2013. Detection of abnormal fish trajectories using a clustering based hierarchical classifier. *In: Proceedings of the British Machine Vision Conference.* British Machine Vision Association.

Bolbol, A., et al., 2012. Inferring hybrid transportation modes from sparse gps data using a moving window svm classification. *Computers, environment and urban systems*, 36 (6), 526–537.

Bolón-Canedo, V., Sánchez-Maroño, N., and Alonso-Betanzos, A., 2015. *Feature selection for high-dimensional data.* Artificial Intelligence: Foundations, Theory, and Algorithms. Cham: Springer International Publishing AG.

Boukerche, A., Zheng, L., and Alfandi, O., 2020. Outlier detection. *ACM Computing Surveys*, 53 (3), 1–37.

Budge, J.S. and Long, D.G., 2018. A comprehensive database for antarctic iceberg tracking using scatterometer data. *IEEE Journal of Selected Topics in Applied Earth Observations and Remote Sensing*, 11 (2), 434–442.

Buja, A., et al., 2009. Statistical inference for exploratory data analysis and model diagnostics. *Philosophical Transactions of the Royal Society A: Mathematical, Physical and Engineering Sciences*, 367 (1906), 4361–4383.

Burges, C.J., 2010. *Dimension reduction: A guided tour.*

Buzin, I.V., et al., 2019. The preliminary results of iceberg drift studies in the russian arctic throughout 2012–2017. *International Journal of Offshore and Polar Engineering*, 29 (4), 391–399.

Cachat, J., et al., 2011. Three-dimensional neurophenotyping of adult zebrafish behavior. *PLoS One*, 6 (3), e17597.

Chan, J., et al., 2021. Small flying object classifications based on trajectories and support vector machines. *Journal of Robotics and Mechatronics*, 33, 329–338.

Chu, R., et al., 2012. A new diverse measure in ensemble learning using unlabeled data. *In: 2012 Fourth International Conference on Computational Intelligence, Communication Systems and Networks*, July. IEEE.

Colefax, A.P., et al., 2020. Assessing white shark (carcharodon carcharias) behavior along coastal beaches for conservation-focused shark mitigation. *Frontiers in Marine Science*, 7.

Collares, L.L., et al., 2018. Iceberg drift and ocean circulation in the northwestern weddell sea, antarctica. *Deep Sea Research Part II: Topical Studies in Oceanography*, 149, 10–24.

Cuadrado-Gallego, J.J. and Demchenko, Y., 2020. *The data science framework: A view from the edison project.* 1st ed. Cham: Springer International Publishing AG.

Da Silveira, N.S., et al., 2016. Effects of land cover on the movement of frugivorous birds in a heterogeneous landscape. *PloS one*, 11 (6), e0156688–e0156688.

De Comité, F., et al., 1999. *Positive and unlabeled examples help learning.* Springer Berlin Heidelberg, 219–230.

Denis, F., Gilleron, R., and Letouzey, F., 2005. Learning from positive and unlabeled examples. *Theoretical Computer Science*, 348 (1), 70–83.

Dodge, S., Weibel, R., and Forootan, E., 2009. Revealing the physics of movement: Comparing the similarity of movement characteristics of different types of moving objects. *Computers, environment and urban systems*, 33 (6), 419–434.

Dodge, S., Weibel, R., and Lautenschütz, A.K., 2008. Towards a taxonomy of movement patterns. *Information visualization*, 7 (3-4), 240–252.

Dosilovic, F.K., Brcic, M., and Hlupic, N., 2018. Explainable artificial intelligence: A survey. *In: 2018 41st International Convention on Information and Communication Technology,*